\documentclass[
reprint,
 amsmath,amssymb,
 aps,
nofootinbib]{revtex4-1}

\usepackage{graphicx}
\usepackage{dcolumn}
\usepackage{bm}
\usepackage{hyperref}
\usepackage{ifthen}
\usepackage{xcolor}

\newcommand{\version}{arXiv}
\renewcommand{\version}{prd} 

\begin{document}


\title{Chiral Magnetohydrodynamic Turbulence}

\author{
  Petar Pavlovi\'c}
\email[]{petar.pavlovic@desy.de}
\author{
  Natacha Leite}
\email[]{natacha.leite@desy.de}
\author{ 
  G\"unter Sigl}
\email[]{guenter.sigl@desy.de}

\affiliation{II. Institute for Theoretical Physics, University of Hamburg,\\
	Luruper Chaussee 149, 22761 Hamburg, Germany }


\begin{abstract}
In this work the influence of the chiral anomaly effect on the evolution of magnetohydrodynamic turbulence was studied. We argue that before the electroweak symmetry breaking and for temperatures high enough such that the electron mass can be ignored, the description of a charged plasma in general 
needs to take into account the interplay between turbulence and the anomaly effects. It was demonstrated that this generalization can have important consequences on
the evolution of turbulence, leading to the creation of maximally-helical fields from initially non-helical ones. Therefore, chiral effects can strongly support turbulent inverse cascade, and lead to a slower decrease of the magnetic field with time, 
and also to a faster growth of the correlation length, when compared to the evolution predicted by the standard magnetohydrodynamics description. Using the weak anomaly approximation, and treating the anomaly contributions
to magnetic energy and helicity as a small perturbation, we derive the specific solutions for the inverse cascade regime that demonstrate how chiral effects support the inverse cascade.

\end{abstract}

\pacs{Valid PACS appear here}
\maketitle

\section{Introduction}
\label{sec:intro}
The phenomenon of turbulence has been confirmed in almost all astrophysical systems, such as solar wind, accretion disks,
galaxy clusters, interstellar medium and  intracluster medium \cite{gold, armstrong, chep, scalo}. Since
astrophysical scales are typically much larger than dissipative scales, these systems are all characterized by high values 
of Reynolds numbers, $Re=Lv/\nu$ (with the characteristic length scale $L$, characteristic velocity $v$ and a kinematic viscosity $\nu$), which is a necessary condition for the establishment of turbulence. Therefore, in general, if the matter of the universe is in a state of movement, turbulence will tend to develop. Since most of the visible matter in the universe is in the state of plasma, characterized by a high conductivity
and permeated by magnetic fields, this turbulence will be described by the set of magnetohydrodynamic (MHD) equations, 
consisting of Maxwell, Navier-Stokes and continuity equation. 

It is generally accepted that magnetic fields 
are present on all scales of the observable
universe \cite{Bernet, Vogt, R.Durrer} and it seems thus natural to assume that they were also present in the early universe. Indeed, a large class of models trying to explain
the observed magnetic fields assumes that they have a cosmological origin \cite{Turner, Bamba, Joyce,Enqvist}. It also seems plausible 
to characterize the early universe by a non-vanishing velocity field, coming from potential first-order phase transitions \cite{kam,kos,tina,chiara,leitao}
or density perturbations \cite{wag}. It then follows that MHD turbulence could be an important phenomenon not only in astrophysical, but also in
the cosmological context. In accord with this reasoning, numerous studies have shown the role that MHD turbulence can have on the evolution
of cosmological magnetic fields and the growth of their correlation length \cite{bran1,bran2, ol,son}.\\ 
For temperatures higher than the temperature of the electroweak transition, the electroweak symmetry is restored and the description
of turbulence should come from considering hypermagnetic ($\mathbf{B}^{Y}$) and hyperelectric ($\mathbf{E}^{Y}$) fields. Hyperfields
introduce the chiral coupling to fermions, coming from the Chern-Simons anomaly in the field Lagrangian. This chiral coupling is
related to the change in the fermion number before the electroweak symmetry breaking, given by
\begin{equation}
\partial_{\mu} j^{\mu} \sim \frac{g'^{2}}{2 \pi^{2}}\mathbf{B}^{Y} \cdot \mathbf{E}^{Y},
\end{equation}
where $g'=e/cos \theta_{W}$, with $\theta_{W}$ being the Weinberg angle. In the case of charged carriers this gives raise to
the effective current which should be added to the standard Maxwell equations. Above the electroweak transition, 
turbulence therefore needs to be properly studied in the context of modified MHD equations, where this effective contribution
and its evolution equation are also considered. Unlike the standard MHD equations, this description leads to the coupling between 
velocity, hyperfields and the particle content of the theory. This framework can thus be important not only for a
better understanding of the evolution of magnetic fields in the early universe, but also from the perspective of various 
baryogenesis models. Baryogenesis and leptogenesis in the context of modified MHD equations around the electroweak transition were extensively studied, but 
all of these contributions ignored the potential role of turbulence \cite{sha, gi, dvo, sem, kamada, kamada2}.\\ 
For temperatures below the electroweak symmetry breaking, where hyperfields are replaced by the ordinary electric and magnetic
fields, the chiral coupling
will not lead to a change in the fermion number, but -- if the lepton mass can be ignored -- the anomaly leads to a change
in the difference between left and right lepton number densities. It was therefore argued that the usual system of MHD equations
should be extended to take into account the effect of the chiral anomaly for high enough temperatures \cite{Kharzeev, Vilenkin, Vilenkin2, yuji, Sidorenko:2016vom, qiu, zadeh}. 
The presence of a chiral asymmetry will lead to an effective electrical current that will also appear in the MHD equations:
\begin{equation}
\mathbf{j_{5}}=-\frac{e^{2}}{2 \pi^{2}}\mu_{5} \mathbf{B}, 
\label{current}
\end{equation}
where $\mathbf{B}$ stands for the magnetic field and $\mu_{5}\equiv (\mu_{L} - \mu_{R})/2$ is the difference between chemical potentials associated with the left- and right-
chiral electrons, respectively. 

Chiral transport phenomena associated with chiral anomaly were recently 
studied theoretically and experimentally in heavy ion collisions \cite{khar, lia}. 
This effect was also explored in the early universe, for temperatures higher than 
$10$ MeV \cite{Boyarsky} and around the electroweak transition \cite{pa}. In the astrophysical context, the suggestion that it may potentially act as a cause of magnetic field enhancement in magnetars \cite{Ohnishi:2014uea, dvosemi1, dvosemi2}, neutron stars \cite{emass, si} and even in quark stars \cite{Dvornikov:2016pai} has been studied, as well as the role it might have in neutrino energy transport in core-colapse supernovae \cite{yam-su}. Some of these objects such as core collapse supernovae, reach high temperatures which makes studies of the chiral magnetic effect in such objects and in the early Universe technically similar.
In \cite{Boyarsky2} the formalism was extended to the case of spatially dependent $\mu_{5}$, and in \cite{gor} it was demonstrated that 
these inhomogeneities do not prevent the anomaly-driven inverse cascade. 
Again, in all the approaches it was assumed that there are no velocity fields and therefore no turbulence occurring, although even some simple estimates
seem to show that it could play a potentially important role both in the early universe \cite{wag, pa} and in neutron stars \cite{si}. Velocity contributions to the MHD equations together with the
chiral anomaly and the chiral vortical effect, in the framework of the early universe, were studied in \cite{hiro}. However the velocity distribution of Ref.~\cite{hiro} was assumed \textit{a priori} to be
given by a standard Kolmogorov spectrum, in general important advection term $\nabla \times (\textbf{v} \times \textbf{B})$ was neglected, and the equation guiding the fluid dynamics was not solved. Recently, scaling laws -- based on the scaling symmetries of the chiral MHD equations -- were
proposed in Ref.~\cite{yam}, but a proper understanding of the chiral MHD turbulence requires further work in the direction of obtaining concrete analytical and numerical
solutions. \\
Motivated by the above arguments, our aim will be to analytically investigate the general interplay between the chiral anomaly and MHD turbulence. We will address this important issue by solving the modified MHD equations in specific limits 
and discuss the general properties of the obtained solutions. 

As it is well known, turbulence remains to be on of the last unsolved problems of classical physics \cite{tu1,tu2}. Due its highly non-linear nature, the Navier-Stokes equation cannot in general be solved analytically to analyze the properties of turbulence. These difficulties become even stronger in the case of MHD turbulence -- leading to still unsolved issues regarding the relationship between magnetic field and velocity field, the role of helicity, proper scaling and time dependence of the quantities of interest,
as well as many other open questions \cite{biskamp,cho}. Even the advanced numerical simulations trying to model MHD turbulence are confronted with difficult challenges and unsolved issues \cite{miesh}. When considering the modification of MHD turbulence by the chiral anomaly effect, which makes the problem even more mathematically difficult, it is not possible to address the issue of chiral MHD turbulence in a simple manner. Therefore, while trying to make the first steps towards a general analytical understanding of chiral MHD turbulence, we will need to consider this problem in specific regimes,
and also use a qualitative reasoning similar to the one typically used in the study of ordinary MHD turbulence. 

This work is organized in the following manner: in \S\ref{sec:ModMHD} the MHD equations in the presence of the chiral anomaly are reviewed and introduced; in \S\ref{sec:invcasc} the behavior of the magnetic helicity is studied; in \S\ref{sec:perturb} we specify regimes for the velocity field and chiral anomaly and obtain solutions to the evolution of fields in the presence of a chiral asymmetry.

\section{Modified MHD equations} \label{sec:ModMHD}

Above the electroweak transition, for high enough temperatures where the chirality flipping processes can be ignored, one can define the chemical potential $\mu_{R}$, associated with the approximately conserved number of right-handed electrons, $n_{R}$. For lower temperatures, but still higher than the
 electroweak scale, one can then perturbatively add the rate of chirality flipping processes to the equations. The number of right-handed electrons is never 
 exactly conserved due to the already mentioned Abelian anomaly. Taking these contributions into account we can write its change in time as \cite{gi} \footnote{Note that different sign conventions have been used in the literature for this relation, but this will be of no interest here, since it does not affect our subsequent analysis.} 
 \begin{equation}      
 \frac{dn_{R}}{dt}=\frac{g'^{2}}{8 \pi^2}\frac{dh^{Y}}{dt} - \Gamma_{s} n_{R},
 \end{equation}
where $h^{Y}=V^{-1} \int \mathbf{A}^{Y} \cdot \mathbf{B}^{Y} d^{3}r$, is the hyperhelicity density, $A^{Y}$ is the vector potential of the hypermagnetic field, and $\Gamma_{s}$ is the chirality flipping rate before the electroweak symmetry is broken.  In the Standard model this leads to a chemical potential for right-handed electrons of the following form \cite{gi} 
\begin{equation}
\frac{d \mu_{R}}{d t}= \frac{1}{T^{2}} \frac{g'^{2}}{8 \pi^{2}}\frac{783}{88}\frac{d h^{Y}}{dt} - \Gamma_{s} \mu_{R} . 
\label{muR}
\end{equation}
On the other hand if the temperature is lower than the electroweak scale -- but high enough such that electron mass can be neglected -- 
one can approximately neglect the difference between the helicity and chirality operators and introduce the density of left/right chiral electrons
and respective chemical potentials, $\mu_{L/R}$.
These temperature conditions are also satisfied in the case of the core of a proto-neutron star.
In this way, the effect of the chiral anomaly leads to the time change of the chiral chemical potential 
given by \cite{Kharzeev}
\begin{equation}
\frac{d \mu_{5}}{d t}=\frac{1}{T^{2}} \frac{3e^2}{4 \pi^2}\frac{d h}{d t} - \Gamma_{b}\mu_{5}, 
\label{mu5}
\end{equation}
where $\Gamma_{b}$ is the rate comprising the chirality flipping processes of the system in question below the electroweak scale, 
$h=V^{-1} \int \mathbf{A} \cdot \mathbf{B} \,d^{3}r$ is the magnetic helicity density, and $A$ is the vector potential. If one considers the change of energy 
associated with this anomaly induced chirality flow, it can in fact be shown that it corresponds to the effective electrical current \eqref{current}. In this work we will assume that fields are slowly varying so that chemical potentials can be treated as
space-independent quantities. 
{This approximation was used in almost all theoretical studies of the chiral magnetic effect, although it is not so straightforward to see how well it describes the systems of interest
and how significant could the effect of chiral chemical potential inhomogeneities actually be. 
It seems that the only way to discuss this issue is to compare the solutions of numerical simulations related to a space-dependent chiral potential to the ones obtained with the
assumption of a space independent chemical potential. This was done in Refs.~\cite{gor,buividovich} and both studies concluded that inhomogeneities of the chiral asymmetry have a negligible 
role for the primordial plasma. Moreover, it was demonstrated in \cite{gor} that the inverse cascade proceeds practically in the same way as in the chirally homogeneous model.
This assumption can therefore be taken as justified, especially since our main concern in this work is the interplay between the anomaly induced and MHD inverse cascade.}

Since the evolution equations have mathematically the same form,
apart from different coefficients and flipping rates, we introduce the following notation to keep the discussion general and 
independent of a specific system 
\begin{equation}
\begin{split}
c_{1}= \frac{g'^{2}}{ \pi^2\sigma} , & \; \; c_{2}=\frac{e^{2}}{4 \pi^2\sigma}  \\  
c_{3}= \frac{g'^{2}}{8 \pi^{2}}\frac{783}{88}, & \; \; c_{4}=\frac{3e^2}{4 \pi^2},
\end{split}
\end{equation}
where $\sigma$ is the conductivity that characterizes the system of interest. Taking into account the contribution of anomaly induced
effective currents to the MHD equations in the resistive approximation \cite{Spruit}, we can write the modified MHD equations in Lorentz-Heaviside units, where we have also used Ohm's law, as
\begin{equation} 
 \nabla \times \textbf{B} = \sigma \left( \textbf{E} - 2 c_{1,2}   \mu_{R,5} \textbf{B} + \textbf{v} \times \textbf{B}\right),
\label{Maxwell}
\end{equation}
\begin{equation}
 \frac{\partial \textbf{B}}{\partial t}= - \nabla \times \textbf{E},
 \label{rot}
\end{equation}
\ifthenelse{\equal{\version}{prd}}
{ \begin{eqnarray}
\label{vel}
 \rho\left[\frac{\partial \textbf{v}}{\partial t} + (\textbf{v} \cdot \nabla) \textbf{v} - \nu \nabla^{2}\textbf{v}\right] &=& \nonumber \\
  - \nabla p + [\sigma \textbf{E} &\times& \textbf{B} + (\textbf{v} \times \textbf{B}) \times \textbf{B}], 
\end{eqnarray} 
} 
{ \begin{equation}
 \rho\left[\frac{\partial \textbf{v}}{\partial t} + (\textbf{v} \cdot \nabla) \textbf{v} - \nu \nabla^{2}\textbf{v}\right]= - \nabla p + \left[\sigma \textbf{E}\times \textbf{B} + (\textbf{v} \times \textbf{B}) \times \textbf{B}\right],
\label{vel}
\end{equation}
} 

\begin{equation}
\frac{\partial \rho}{\partial t}+ \nabla(\rho \cdot \textbf{v})=0,
\label{c}
\end{equation}
\begin{equation}
\frac{d \mu_{R, 5}}{d t}=\frac{1}{T^{2}} c_{3,4} \frac{d h}{d t} - \Gamma_{f} \mu_{R, 5}+ \Pi_{sr}, 
\label{mutot}
\end{equation}
where  $\rho$ is the matter density, $ \Gamma_{f}$ is the total chirality flipping rate, and we have also added a possible source term, $ \Pi_{sr}$,
to take into account possible processes which generate a chiral asymmetry $\mu_{R, 5}$ in a given system. 
We note that, for instance, in the core of a neutron star it follows $\Pi_{sr}= \Gamma_f \mu_{R,5}^{b}$, where $\mu_{R,5}^{b}$ is the equilibrium value
of the chiral potential of the background medium in the absence of magnetic helicity \cite{si}. $\textbf{E}$ and $\textbf{B}$ denote from hereon both hyper and ordinary electric and magnetic fields, respectively. 

This approximation of the MHD equations assumes high conductivity, as well as the global neutrality of plasma, i.e. $\nabla \cdot \textbf{J}=0$ and $\nabla \cdot\textbf{E}=0$, and the displacement current is neglected. 
These equations will have the same form on curved spacetime, which is of interest 
in the cosmological context, as long as time is replaced by conformal time and all the physical quantities
are scaled with the conformal factor \cite{Subramanian2, Banarjee}.
In this work we will concentrate on the case of the incompressible fluid, where the continuity equation reduces to the condition $\nabla \cdot \mathbf{v}=0$.  This condition will
be physically satisfied if the ratio between the fluid velocity and the speed of sound in the fluid is much smaller than unity, and we will moreover assume that the bulk flow velocity is 
non-relativistic. 
This in fact needs to be the case for velocity fields associated to magnetic fields of realistic cosmological strengths, such that they do not come into contradiction with the 
established course of primordial nucleosynthesis and microwave background fluctuations. 
For instance, as discussed in Refs.~\cite{zeldovich,grishchuk,gasperini},  cosmological magnetic fields act as a source of cosmic microwave background (CMB) and create characteristic anisotropy patterns, thus leading to constraints on their magnitude in order to be consistent with the CMB observations \cite{shiraishi}. Additionally, magnetic fields present at the time of primordial nucleosynthesis can influence the formation of light nuclei, for example by enhancing the rate of the expansion of the Universe \cite{greenstein,cheng,grasso}, which again bounds their possible values to the ones consistent with highly non-relativistic bulk velocities. 
For the case of the early Universe it can easily be shown that the incompressible 
MHD fluid is a good approximation \cite{subra}. As it is well known, fluids can be treated as incompressible if there are no pressure variations which would cause a change in density.
 For the sake of an estimate, taking the fields frozen in the expanding Universe, the ratio between magnetic pressure and fluid radiation pressure is typically
$B^{2}/8 (\pi p) \approx 10^{-7}$. Since radiation pressure is homogeneous and isotropic, and magnetic pressure is negligible with respect to it,
pressure variations in the fluid can be neglected, $B^{2}/8 (\pi p) \ll 1$, and the early Universe plasma can be treated as an incompressible fluid to a very good approximation. 

One can easily see that if there are no initial chiral asymmetry and magnetic helicity present, then the system described by the equations \eqref{Maxwell}-\eqref{mutot} will just evolve according to
the standard decaying MHD turbulence description. 
On the other hand, if the fields are initially helical 
with initially vanishing chiral asymmetry, a finite $\mu_{5}$ will in general be generated (and vice versa) \cite{pa}, which implies that the evolution for later times will be different from the standard MHD description. 

Apart from the already mentioned kinetic Reynolds number, $Re$, it is also useful to define the magnetic Reynolds number $Re_{B}\equiv \ 4 \pi Lv \sigma$. A turbulent MHD regime typically corresponds to $Re\gg 1$ and $Re_{B}\gg 1$.

The analysis of equations \eqref{Maxwell}-\eqref{mutot} is mathematically simpler decomposing the (hyper)fields into Fourier components
\begin{equation}
\mathbf{B}(\mathbf{r},t)=\int \frac{d^{3} k}{(2 \pi)^{3}}e^{i\mathbf{k} \cdot{\mathbf{r}}}\mathbf{B}(\mathbf{k},t) 
\end{equation}
Focusing our attention on the evolution of statistically homogeneous and isotropic magnetic fields we obtain the condition \cite{monin}
\ifthenelse{\equal{\version}{prd}}
{ \begin{eqnarray}
\label{correlator}
\langle B_{i}(\mathbf{k},t) && B_{j}(\mathbf{q},t)\rangle = \frac{(2 \pi)^3}{2} \delta(\mathbf{k} +\mathbf{q})\nonumber \\
&&\left[(\delta_{ij}-\hat{k}_{i} \hat{k}_{j})S(k,t)
+ i \epsilon_{ijk}\hat{k}_{k}A(k,t)\right], 
\end{eqnarray} 

} 
{ \begin{equation}
\label{correlator}
\langle B_{i}(\mathbf{k},t) B_{j}(\mathbf{q},t)\rangle = \frac{(2 \pi)^3}{2} \delta(\mathbf{k} +\mathbf{q})
\left[(\delta_{ij}-\hat{k}_{i} \hat{k}_{j})S(k,t)
+ i \epsilon_{ijk}\hat{k}_{k}A(k,t)\right], 
\end{equation} 

} 

where $\hat{k}_{i}$ is the unit vector of $\mathbf{k}$, and $S(k,t)$ and  $A(k,t)$ denote the symmetric and antisymmetric parts
of the correlator, respectively. Using \eqref{correlator} we can write the magnetic energy density, $\rho_{m}$ and helicity density, $h$, in the volume $V$ as
\begin{equation} \label{eq:men_den}
\rho_{m}=\frac{1}{2V} \int d^{3} r\langle \mathbf{B}^{2}(\mathbf{r},t)\rangle = \int d\ln k \, \rho_{k}(t),
\end{equation}
\begin{equation} \label{eq:hel_den}
 h=\frac{1}{V} \int d^{3} r\langle \mathbf{A}(\mathbf{r},t) \cdot \mathbf{B}(\mathbf{r},t)\rangle=\int d\ln k \, h_{k}(t),
\end{equation}

where we have introduced the spectral magnetic energy and helicity $\rho_{k}(t)=k^{3}S(k,t)/(2 \pi)^{2}$ and $h_{k}(t)=k^2 A(k,t)/2 \pi^{2}$, respectively. The maximal value for helicity density is achieved if all the magnetic energy is stored in one circularly polarized mode \cite{hiro}, and thus 
$\rho_{k}(t)=(k/2)h_{k}(t)$. This configuration of magnetic field is called maximally helical.

As usual, the fluid part of turbulence is characterized by the kinetic energy, $\rho_{K}=(1/2V)\int d^{3} r \rho  v^{2} $. The relative importance of kinetic over magnetic effects in turbulence will be measured by the ratio between the respective
energy densities, $\Gamma=\rho_{K}/ \rho_{m}$, which will in general be a function of time. 
Turbulence will develop on scales between the dissipation scale, where the Reynolds number becomes small and turbulence stops due to dissipation processes, and the scale of the largest magnetic eddies. The latter  is modeled by the magnetic correlation length
\begin{equation}
\xi_{m}=\frac{\int   k^{-1} \rho_{k} d \ln k}{\rho_{m}}, 
\label{corr}
\end{equation}
and the kinetic correlation length can be defined in a similar fashion. Non-linear turbulent phenomena are dominant in the inertial interval -- the interval between the scales where injection and dissipation effects become relevant.

In order to study the evolution of magnetic energy and helicity in different regimes, let us write the evolution of the magnetic field, described by \eqref{Maxwell} and \eqref{rot}, in Fourier space  
\ifthenelse{\equal{\version}{prd}}
{ \begin{eqnarray}
\label{eq:B_ev}
 \partial_t \mathbf{B_k}=&& - \frac{k^2}{\sigma}  \mathbf{B}_\mathbf{k} - 2 c_{1,2} \mu_5 (i\mathbf{k}\times \mathbf{B}_\mathbf{k}) \nonumber \\
 &&+ \frac{i}{(2\pi)^{3/2}}\mathbf{k}\times\int d^3q (\mathbf{v}_\mathbf{k-q}\times \mathbf{B}_\mathbf{q}) \, ,
\end{eqnarray}
} 
{ \begin{equation}
\label{eq:B_ev}
 \partial_t \mathbf{B_k}= - \frac{k^2}{\sigma}  \mathbf{B}_\mathbf{k} - 2 c_{1,2} \mu_5 (i\mathbf{k}\times \mathbf{B}_\mathbf{k}) + \frac{i}{(2\pi)^{3/2}}\mathbf{k}\times\int d^3q (\mathbf{v}_\mathbf{k-q}\times \mathbf{B}_\mathbf{q}) \, ,
\end{equation}    
} 

where $\mathbf{B}_\mathbf{k}\equiv\mathbf{B}(\mathbf{k},t) $ and $\mathbf{v}_\mathbf{k-q}\equiv\mathbf{v}(\mathbf{k - q},t) $. From here on we denote  the integral term containing the velocity field as
\begin{eqnarray}
\mathbf{I_k} = \frac{i }{(2\pi)^{3/2}} \int d^3q (\mathbf{v}_\mathbf{k-q}\times \mathbf{B}_\mathbf{q}) \, 
\end{eqnarray}

The time evolution of the power spectra $\rho_k$ and $h_k$ defined in \eqref{eq:rhok_ev} and \eqref{eq:hk_ev} can then be derived from the magnetic field evolution by multiplying \eqref{eq:B_ev} and its the complex conjugate by $\mathbf{B}_\mathbf{k}^*$ and $\mathbf{B}_\mathbf{k}$, respectively. Analogously, the evolution of \eqref{eq:hel_den} is obtained by multiplying \eqref{eq:B_ev} by the vector potential complex conjugate. This leads to the following expressions for the spectral evolution
\begin{equation} \label{eq:rhok_ev}
\partial_t  \rho_k= - \frac{2 k^2}{\sigma} \rho_k -  c_{1,2} \mu_5 k^2 h_k + I_{1}(k)\, ,
\end{equation}  
\begin{equation} \label{eq:hk_ev}
 \partial_t  h_k= - \frac{2 k^2}{\sigma} h_k - 4 c_{1,2} \mu_5 \rho_k + I_{2}(k) \, ,
\end{equation}  
where $I_{1}(k)=  k \left[(\mathbf{k}\times \mathbf{I_k}) \cdot \mathbf{B_{k}}^{*}+(\mathbf{k}\times \mathbf{I_k}^{*}) \cdot \mathbf{B_{k}} \right] $ and $I_{2}(k)= k \left[(-i) \mathbf{I_k} \cdot \mathbf{B_k}^{*} + \mathbf{A_{k}} \cdot (\mathbf{k} \times \mathbf{I_k}^{*})\right]$.

\section{Inverse cascade and the role of helicity} \label{sec:invcasc}

Magnetic helicity, which measures the global topology of field lines by describing their linking and twisting, is an important 
quantity for the analysis of different MHD flow structures, and it is known to be conserved in ideal MHD, i.e. when $\sigma \rightarrow \infty$. 
Conservation of helicity can also be shown in the chiral case. 
Starting from the definition of helicity density, as in \eqref{eq:hel_den}, one has $\dot{h}=(1/V)\partial_{t}\int d^3 r\, \mathbf{A} \cdot \mathbf{B}$, and with \eqref{Maxwell} and 
\eqref{rot}, it follows
\begin{equation} \label{eq:hel_ev}
\frac{dh}{dt}=-\frac{2}{V} \int d^3{r} \left[\frac{1}{ \sigma}\left(\nabla \times \textbf{B}\right)\cdot  \textbf{B} +2c_{1,2} \mu_{5,R}|\textbf{B}|^{2}\right]. 
\end{equation}
Using for  the current $\mathbf{J}=\nabla \times \textbf{B}$ and introducing the effective chiral current $\mathbf{J}_{R, 5}=2 \sigma c_{1,2} \mu_{R,5}  \mathbf{B}$,
Eq.~\eqref{eq:hel_ev} can be put in a form that resembles the corresponding equation for standard MHD: $ \dot{h}=-2/(V \sigma) \int d^3{r}
(\mathbf{J}+ \mathbf{J}_{R, 5}) \cdot \mathbf{B}$. This means that in the limit $\sigma \rightarrow \infty $, there can be no change of $\mu_{5}$ due to the chiral anomaly, even if one has an initially present chiral asymmetry.

Similarly, using once more Eqs. \eqref{Maxwell} and 
\eqref{rot}, and applying the Stokes theorem, we obtain that the change of magnetic flux, $\Phi= \int_{S} \mathbf{B} \cdot d\mathbf{S}$,
is given by
\begin{equation}
\frac{d \Phi}{dt}=-\frac{1}{ \sigma} \oint_{\ell} \mathbf{J} \cdot d\mathbf{\ell} - 2c_{1,2} \mu_{R, 5}\oint_{\ell} \mathbf{B} \cdot d\mathbf{\ell} . 
\end{equation}
Therefore, in the case of ideal chiral MHD, the magnetic flux will also be conserved, while in the case of finite conductivity, the flux changes -- corresponding to cutting and reconnecting field lines (that are no longer frozen in the plasma) -- will be enhanced by the chiral anomaly.

Apart from  the well known hydrodynamical direct cascade -- energy transfer from large to small scales --MHD turbulence
can also undergo an inverse cascade. 
This energy transport of a conserved quantity from small to large scales represents an important process
of self-organization of turbulent structures (effectively measured by the correlation length), therefore leading to the development of order from initial chaotic conditions \cite{po}. 
Helicity is known to play an important role in the establishment 
of inverse cascades. Since, for high conductivities, helicity is a quasi-conserved quantity,
short-scale modes cannot be significantly washed out, and their magnetic helicity gets transferred to large scale modes. But whether the presence of helicity is a necessary
condition for the development of inverse cascade is still not completely resolved. While some previous analytical \cite{son, camp} and numerical \cite{christenson, ban-jed} analysis 
concluded that there is no inverse cascade for non-helical fields, some recent findings \cite{zrake, tev} seem to show that
inverse cascades may be possible even if the fields are non-helical. It was also argued that non-helical inverse cascades
can exist due to scaling symmetries of the MHD equations \cite{ol}. One of the difficulties with this issue is that in principle it may
be hard to distinguish between inverse cascade and resistive damping (discussed by Son in Ref.~\cite{son}) especially since helicity is not
an exactly conserved quantity for finite conductivities. 
In any case, it seems obvious that the presence or absence of helicity will have an important
impact on the evolution of MHD turbulence since it will strongly support an inverse cascade -- even if it is not a necessary condition for it. 
It is essentially at this point that the chiral anomaly effect can lead to important changes in the turbulent MHD evolution, due to its
property of creating helical magnetic fields from non-helical ones.

We give here a general analytical discussion on the influence of the anomaly
on MHD turbulence in the following manner. Let us initially, starting from a time $t_{0}$, consider only the non-helical MHD turbulence. According to the analytical results
in Ref. \cite{Olesen3}, which seem to be consistent with the result of numerical simulations in Refs. \cite{zrake, tev}, in this regime we consider that the magnetic energy density scales like
\begin{equation}
\rho_{k}(t)=\sqrt{\frac{t_{0}}{t}}k \rho_{k}\left(k\sqrt{\frac{t}{t_{0}}},t_{0}\right) ,
\label{nohel}
\end{equation}
such that the total magnetic energy density, $\rho_{m}$, scales like $\rho_{m} \sim 1/t$. This scaling can also be 
introduced using simple analytical arguments, as done in Ref. \cite{biskamp}. Using \eqref{corr}, we see that the correlation length
then grows as $\xi_{m} \sim \sqrt{t}$.
If now, at some later time $t_{i}$, a finite $\mu_{5}$ is created due to some particle processes, it will then, according to \eqref{eq:hk_ev}, lead to a change in helicity. After some short time interval, $\Delta t \equiv t -t_{i}$ there will be a finite helicity created according to
\begin{equation}
h_{k}=-4c_{1,2} \int_{t_{i}}^{t} \mu_{5}(t) \sqrt{\frac{t_{0}}{\tau}} k \rho_{k}\left(k\sqrt{\frac{\tau}{t_{0}}},t_{0}\right) d \tau  + \int_{t_{i}}^{t} I_{2} d \tau \, .
\label{helind}
\end{equation}
Concentrating now on the case where $ \Gamma <1$, so that the second integral can be neglected, we assume that $\mu_{5}$(t) is a smooth function and can therefore be written as $\mu_{5}(t)= \sum_{n} c_{n}t^{n}$ on a small time interval $\Delta t$. Here the values of the expansion coefficients, $c_{n}$ are determined
by the concrete form of the flipping and source terms entering in \eqref{mutot}, which is different for different systems of interest. The total anomaly induced helicity density is given by
\begin{eqnarray} \label{eq:hinduced}
h_{in}(t) &=& \int h_{k} d\ln k \nonumber \\
&=& - 4 c_{1,2} t_{0} \sum_{n} c_{n} \frac{(t^n - t_{i}^n)}{n} \int_{0}^{\infty} \rho_{k}(x, t_{0})dx,
\end{eqnarray}
so that the time evolution of the induced helicity, $h(t) \sim t^{n}$, is determined by the evolution of $\mu_{5}(t)$. 

On the other hand, in the special case of $\mu_{5}=const.$ one gets a logarithmical scaling with time
\begin{equation}
 h_{in}^{sta}= -4 K c_{1,2}t_{0}  \log\left( \frac{t}{t_{i}}\right) \int_{0}^{\infty} \rho_{k}(x,t_{0})dx .
\end{equation}
The helicity density growth will approximately follow $h_{in} \sim t^{n} $  or $h_{in}^{sta} \sim \log(t/t_{i}) $ as long as the term in Eqs. \eqref{eq:hk_ev} and \eqref{eq:rhok_ev} containing 
$h_{k}$ remains much smaller than the term containing the energy density. This signifies the conclusion that the creation of helicity in a initially non-helical turbulent plasma due to the chiral anomaly effect is a transitory phenomenon. This is again in accord with considering the induction
of helicity on a small time interval $\Delta t$.
When the induced helicity reaches a level where the first term on the r.h.s. of \eqref{eq:hk_ev} becomes comparable
   to the second term proportional to the magnetic energy term, the above approximation can no longer be applied
and one needs to consider the full set of coupled differential equations for energy end helicity density. In that regime,
even a qualitative understanding of the interrelationship between MHD turbulence and the chiral anomaly effect is not so 
simple, since on the one hand chiral anomaly also leads to inverse cascade \cite{Boyarsky,pa}, but on the other hand
MHD inverse cascade is supported by the conservation of helicity, while the anomaly effect is based on the change of helicity, 
as discussed above. 
It therefore seems natural that a system will tend to approach the state where one of the effects -- either MHD turbulence or the chiral anomaly -- dominates and determines the main features of its dynamics, while the other one has a minor role, which can be treated as a correction.

We can first concentrate on the case where anomaly effects remain so small that the only significant contribution from the chiral 
anomaly is the afore discussed creation of helicity, after which helical magnetic fields follow essentially the same evolution as in standard
MHD. 
We expect this to happen in systems characterized by high conductivities, suppressing the change of total helicity,
and when the chirality flipping rates are strong compared to the source term for $\mu_{5}$. 
We will here generalize the usual scaling arguments, used in the study of MHD and discussed in standard textbooks \cite{biskamp}, to the anomalous case. As discussed in \cite{biskamp}
such arguments, although approximate in nature, lead to a satisfactory matching with the results of advanced numerical simulations -- so it is justified to apply them to the chiral MHD case.
In this regime we have the approximated scaling 
\begin{equation}
\rho_{m} \xi_{m} \sim \frac{h_{in}}{2} \approx const.
\label{magintsc}
\end{equation}
coming from the fact that helicity -- determined by correlation length and magnetic energy according to \eqref{corr} -- will be approximately conserved during the developed turbulence leading to inverse cascade. 
Using a Kolmogorov-like reasoning, i. e. assuming a constant energy transfer rate proportional to the eddy-turnover rate in the inertial interval, where the dissipation effects can be neglected, we can write
\begin{equation}
\frac{d \rho_{tot}}{dt} \sim - \rho_{tot} \frac{ \rho_{K}^{1/2}}{\xi_{m}},
\end{equation}
where $\rho_{tot}=\rho_{m} + \rho_{K}$. 
This can be further expressed as
\begin{equation} \label{eq:scaling}
\frac{d}{dt}\left[\rho_{m}(1+ \Gamma)\right] \sim - \frac{\rho_{m}^{5/2}\Gamma^{1/2}(1+ \Gamma)}{h_{in}}. 
\end{equation}
Numerical simulations typically show that the ratio between kinetic and magnetic energy asymptotically approaches a constant value in the case of standard MHD turbulence \cite{ban-jed}. It is clear that this will remain
so in the regime where the anomaly effects are small compared to the standard turbulent MHD background. Using this fact, $\Gamma$ can be treated as independent of time, and then from \eqref{eq:scaling} it follows that the resulting scaling will be
$\rho_{m} \sim t^{-2/3}$. 
Therefore, even the presence of a weak chiral anomaly effect in initially non-helical MHD 
turbulence will tend to change the time evolution of the magnetic energy from $\rho_{m} \sim 1/t$ to $\rho_{m} \sim t^{-2/3}$.
This comes as a result of the anomaly induced helicity according to (\ref{helind}), which then plays the role of a (quasi-)conserved
quantity, leading to a slower decrease of the magnetic energy, to a faster growth of the correlation length and additionaly supports
the inverse cascade. 
If one would use the scaling solution proposed in \cite{Banarjee}, a causal tail at large scales, $l=5$, corresponds to $\rho_m \propto t^{-10/7}$ and $\xi_m \propto t^{2/7}$, then the difference induced by the chiral anomaly effect would be even larger.  

At later times in the system's evolution, this initial anomaly induced helicity, $h_{in}$ will then lead to the realization
of maximally helical fields -- since magnetic fields with initial fractional helicity become maximally helical 
due to standard MHD turbulence \cite{andrey, campanelli2}. We thus conclude that the chiral MHD turbulence, when anomaly effects are small compared
to the velocity and magnetic field terms, will tend to create maximally helical fields from non-helical fields. 

When this 
maximally helical state is reached, then using \eqref{corr}, we have $\rho  \xi_{m} \approx h/2 \approx const.$, which implies $\ \xi_{m} \sim t^{2/3}$. This represents a growth of organized turbulent structures faster than the already discussed scaling associated with \eqref{nohel}.

Turning to the other regime, where chiral anomaly effects cannot be neglected after the helicity was induced, we assume that 
all violation to helicity conservation comes from the time change of the anomalous chemical potential: $\dot{h} \sim \kappa_{T} \dot{\mu}_{5}$, where $\kappa_{T}=T^{2}/c_{3,4}$. 
Such is the case
when the chirality flipping and source rates are either both small or compensate each other. When this is not the case, we can use the same logic that will be presented, but taking the complete
Eq. \eqref{mutot}. 
It is still possible to define a magnetic integral scale, as in \eqref{magintsc}, at a given instance in time, but
this scale will be time dependent. For simplicity, we 
will here consider only the cases where the variation of the temperature can be neglected, but this treatment can be easily extended. 
It now follows
\begin{equation}
\frac{d}{dt}\left[\rho_{m}(1+ \Gamma)\right] \sim - \frac{\rho_{m}^{5/2}\Gamma^{1/2}(1+ \Gamma)}{\kappa_{T} \mu_{5} } \, . 
\end{equation} 
Considering again that $\Gamma$ approaches a constant value, $\mu_{5}(t)= \sum_{n} c_{n}t^{n}$, we obtain 
\begin{equation}
\rho_{m} \sim \left(\int\frac{dt}{k_{T}\sum_n c_{n} t^{n}}\right)^{-2/3} .
\end{equation}
Focusing on the special case where $\Gamma_{f}$ and $\Pi_{sr}$ are such that $\mu_{5}$ can be approximated by a power-law solution, $\mu_{5}=Kt^{n}$, which is of practical interest in several contexts,  
we obtain $\rho_{m} \sim  t^{2(n-1)/3}$.
If the maximally helical regime is reached, then
$\xi_{m} \approx  \kappa_{T} \mu_{5}/(2 \rho_{B}) \approx t^{(n+2)/3} $. 
We can therefore clearly see that, in this regime, the evolution of the chiral asymmetry is governing the overall evolution of the magnetic field and correlation length. It also directly follows that for a sufficiently fast growth of $\mu_{5}$, with $n>1$, the total magnetic energy will grow in time. In this case, the system transforms the energy stored in the chiral asymmetry chemical potential into magnetic field to such an extent that it completely changes the dynamics of MHD turbulence.
As previously discussed in Ref. \cite{hiro}, when the
advection term in the MHD equations is discarded, $\mu_{5}$ will have an attractor solution with $n=-1/2$. Taking this specific value, we get 
$\xi_{m} \sim t^{1/2}$. For this regime, we have hence independently confirmed the scaling laws recently proposed in 
\cite{yam}, which were there derived using symmetry arguments.\\
In the case where chirality flipping and source rates are neither small nor compensate each other, one in general needs to consider their contribution to the change of helicity of a given system. Then the scaling of magnetic energy with time
will be given by the solutions of the following equation, determined by the specific form of $\Gamma_{f}$ and $\Pi_{sr}$
\begin{equation}
\frac{d\rho_{m} }{dt} \sim - \frac{\rho_{m}^{5/2}}{k_{T}\sum c_{n} t^{n} - \int\left(\Pi_{sr} - \Gamma_{f} \sum_n c_{n} t^{n} \right)dt}\, .
\end{equation}
Finally, we discuss the case where initially helical MHD turbulence reached inverse cascade regime with $\mu_{5}=0$ and then, at time $t_{i}$, a finite chiral asymmetry is
generated due to to the particle processes. We will argue that chiral turbulence will then in general tend to develop to either anomaly dominated or turbulence dominated regime. 
In the inverse cascade regime $h \approx const$ and therefore ${d\mu_{R,5}}/{dt}= -\Gamma_{f}\mu_{R,5}+ \Pi_{sr}$. In the case of  $\Pi_{sr} < \Gamma_{f}\mu_{R,5}$ asymmetry
will be washed out fastly and system will evolve according to the standard MHD picture. On the other hand, when  $\Pi_{sr} > \Gamma_{f}\mu_{R,5}$ the chiral asymmetry chemical
potential will grow until it becomes significant enough that helicity can no longer be treated as constant and system exits turbulent inverse cascade regime. 
So, depending on the relative strength of source and flipping term, turbulence will tend to support or wash out the anomaly, as a result of helicity conservation in the inverse
cascade regime.

\section{Weak anomaly regime} \label{sec:perturb}

In the last section we have considered the regime of chiral MHD turbulence with no initial helicity, as well as the weak and strong anomaly regimes, in general -- based
on qualitative arguments.
We next turn to the weak anomaly regime of the chiral MHD turbulence in more detail, i.e. when $ |2 c_{1,2}  \mu_{R, 5} \textbf{B}| \ll |\textbf{v} \times \textbf{B}+\textbf{E}|$, with the aim of obtaining
concrete solutions for the evolution of magnetic fields. 
As discussed earlier, the non-linear nature of the Navier-Stokes equation, together with the coupling between velocity and magnetic field, makes even the general analytical study of the non-chiral MHD equations to still remain as an unsolved issue, and numerical simulations to be a highly non-trivial task. Is is therefore reasonable that the first step towards an understanding of the chiral MHD turbulence should be the consideration of simplified regimes, such as the described weak anomaly regime. It is then possible to use the already known properties of standard MHD turbulence and observe the modifications induced
by the chiral anomaly effect. The weak anomaly regime is also of physical interest in at least some important cases, for instance around the electroweak transition where the chiral effects
are expected to be small with respect to the standard MHD background \cite{pa}. 
We expect that the overall dynamics will in this regime be determined by the usual MHD terms (i.e. $\mu_{R,5}=0$ case), and that anomaly effects will have the role
of a correction to these results. It is reasonable to assume that in this case the
time scale will also be determined by purely MHD considerations. Namely, one can introduce typical time scales: Alfv\'en time scale, $ \tau_{A}= l_{\parallel}/v_{A}$, eddy time scale, $\tau_{s}= l_{\perp}/v_{l}$ and anomaly growth time scale,
$\tau_{g}=1/\Gamma_{g}$ . Here $\Gamma_{g}=c_{1,2} \mu_{5} k_{5}/4$ \cite{pa}, with $k_{5}=c_{1,2} \sigma \mu_{R,5}$, $v_{A}=B/\sqrt{ \rho}$, and $l_{\parallel}$ and $l_{\perp}$ are the length intervals in the parallel and perpendicular directions to 
the magnetic field. Thus, in this regime it follows $ \tau_{g} \gg \tau_{A} \sim \tau_{s}$ where the last equality
between the Alfv\'en time scale and eddy time scale comes as a result of the critical balance condition in the Goldreich-Sridhar model of standard MHD turbulence \cite{srid}. 
We can then treat the anomaly contribution to magnetic energy and helicity as a small perturbation to the standard background MHD solutions, which we label as
 $\rho^{\rm bg}$ and $h^{\rm bg}$. We then have $\rho_{B}=\rho^{\rm bg} + \rho^{\mu} $, $h=h^{\rm bg} + h^{\mu}$. 
We now add this perturbation to the equations \eqref{eq:rhok_ev} - \eqref{eq:hk_ev} around $\rho_{B}=\rho^{\rm bg}$ and $\mu_{5}=0$ and ignore 
all the terms higher than the first order in perturbation. Thus, to zeroth order one obtains:
\begin{equation}
\partial_t  \rho^{\rm bg}_k= - \frac{2 k^2}{\sigma} \rho^{\rm bg}_k  + I_{1}(k)\, ,  
\end{equation}
\begin{equation}
\partial_t h^{\rm bg}_k= - \frac{2 k^2}{\sigma}  h^{\rm bg}_k + I_{2}(k) ,
\end{equation}
and to first order:
\begin{equation}
\partial_t  \rho^{\mu}_k= - \frac{2 k^2}{\sigma} \rho^\mu_{k} -  c_{1,2} \mu_5 k^2 h^{\rm bg}_k \, ,
\end{equation}
\begin{equation}
\partial_t  h^\mu_{k} = - \frac{2 k^2}{\sigma} h^\mu_{k} - 4 c_{1,2} \mu_5 \rho^{\rm bg}_k \, ,
\end{equation}
where we assumed that the coupling between the chiral asymmetry, $\mu_{5}$, and velocity can be neglected. This will for instance be the case
for systems where $\Gamma <1$, such that velocity contributions to the perturbation equation can be neglected. Henceforward we focus on this regime. Both equations above, for $\rho^\mu_{k}$ and $h^\mu_{k}$, have a general analytical solution,
namely
\begin{equation}
\rho^\mu_{k}= \frac{\int e^{\frac{2 k^{2}}{\sigma}t} f(t) dt}{e^{\frac{2 k^{2}}{\sigma}t}} + const.
\label{pert1}
\end{equation}
and
\begin{equation}
h^\mu_{k}=\frac{\int e^{\frac{2 k^{2}}{\sigma}t} g(t) dt}{e^{\frac{2 k^{2}}{\sigma}t}} + const. \, , 
\label{pert2}
\end{equation}
where $f(t)= - c_{1,2} \mu_5 k^2 h^{\rm bg}_k $ and $g(t)=- 4 c_{1,2} \mu_5 \rho^{\rm bg}_k$. 

\begin{figure}
\includegraphics[width=0.42\textwidth]{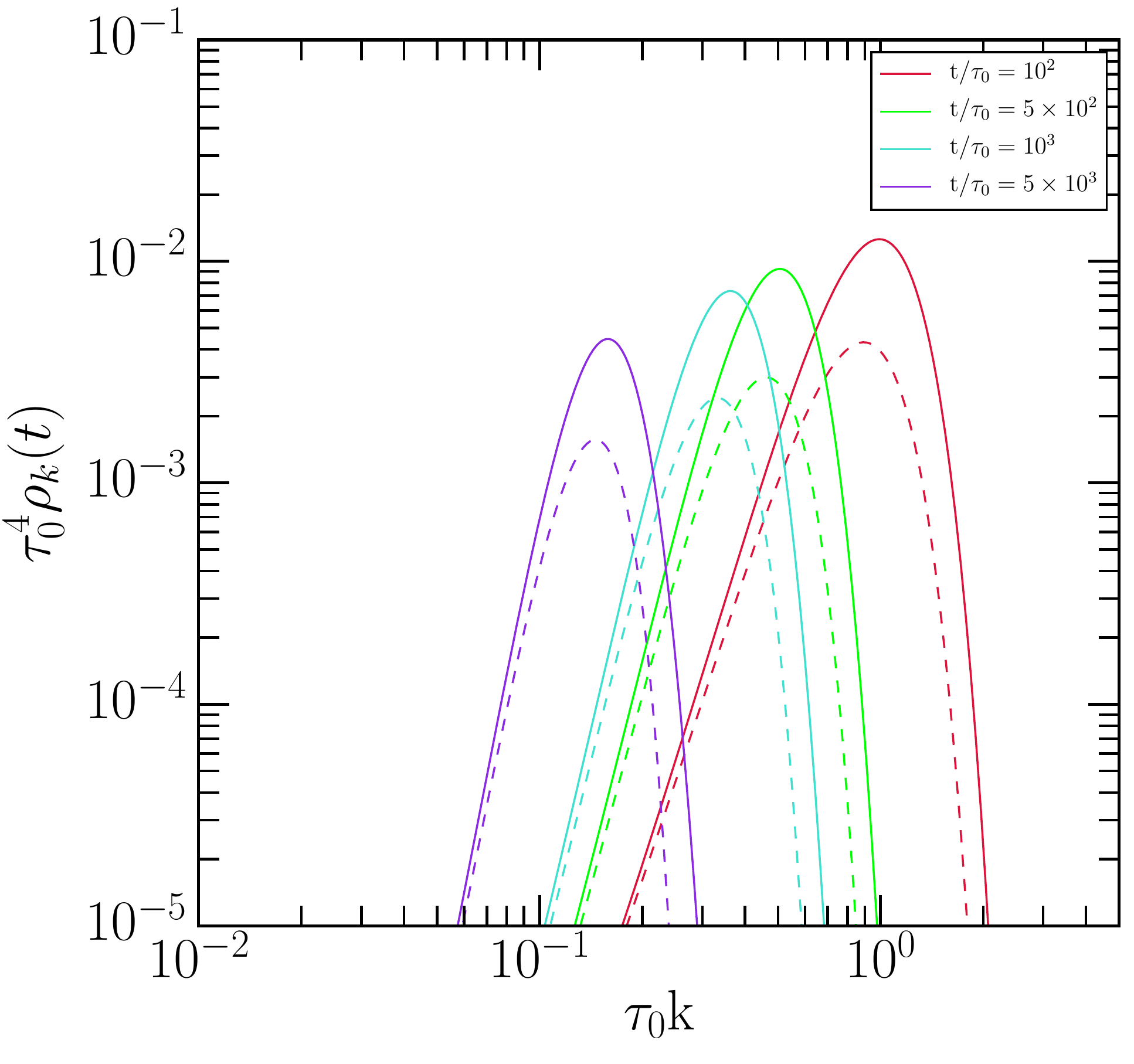}
\includegraphics[width=0.42\textwidth]{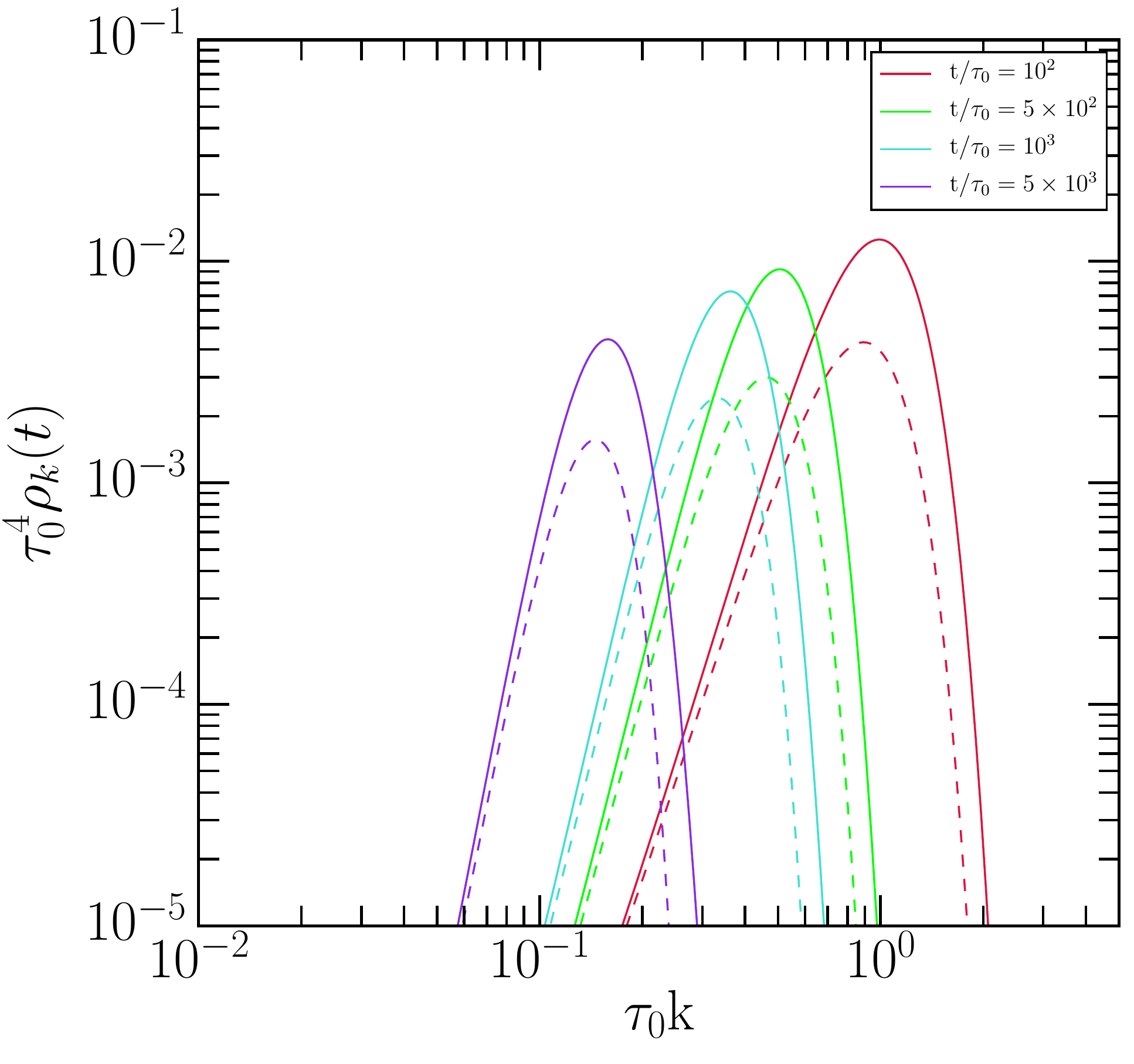}
\includegraphics[width=0.42\textwidth]{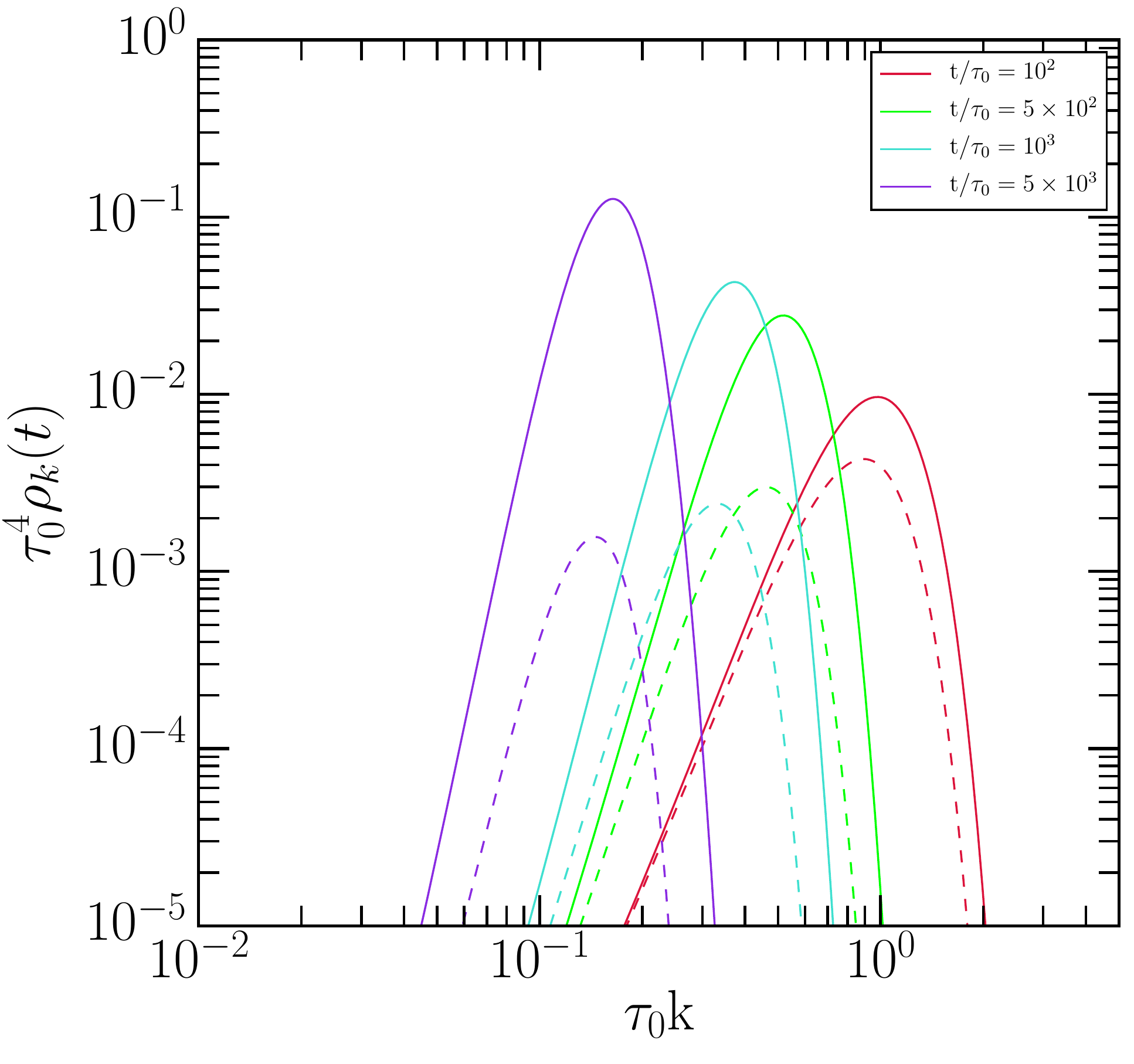}
\caption{\label{pic:3hels} Total magnetic energy density spectrum (solid) and background magnetic energy density spectrum (dashed) at different times, computed from \eqref{pert1} and \eqref{mutot}, and \eqref{background1}, respectively, with $\mu_5^0=-10^3\tau_0^{-1}$ and $\epsilon=7\times 10^{-3}$. \textit{Upper panel:} For $\Gamma_f\mu_5 = \Pi_{sr}$; \textit{Central panel:} For $\Gamma_{f}\propto t^{-2}\tau_0^{-1}$ and $\Pi_{sr}=0$.
\textit{Lower panel}: For $\Gamma_{f}=0$ and $\Pi_{sr}=const.$}
\end{figure}

In order to proceed
we need some analytical model for the background solutions, that is, for the standard MHD turbulence. We will therefore 
follow the approach proposed in Refs.~\cite{camp, campanelli2}.
Since we are interested in the evolution of turbulence
in the inertial interval, far enough from the dissipation scale, we can neglect the dissipation term $\nu \nabla^{2}\textbf{v}$
in Eq. \eqref{vel}. Moreover, following the approach of Ref.~\cite{cor} and Ref.~\cite{camp}, we assume that the Navier-Stokes equation \eqref{vel} can be
quasi-linearized neglecting the term $(\textbf{v} \cdot \nabla) \textbf{v}$, which is justified as long as the ratio between 
the fluctuating and average part of the velocity is much smaller than $\Gamma$. 
This is also the case for systems where the velocity is small, $\Gamma<1$, such that the magnetic effects dominate over the kinetic effects.
In general, the exact values of the characteristic velocity and field scales
are not well known -- especially in the early universe -- and estimates are strongly dependent on the concrete models. In any case, it is reasonable to expect that this regime will be reached
in different cosmological and astrophysical contexts.
The next assumption is that the Lorentz force per volume, $\mathbf{F}_L = \mathbf{J}\times \mathbf{B}$, is the responsible mechanism for turbulence to occur and we therefore take $\partial_t \mathbf{v}\approx \mathbf{F}_L$ on large scales.
We stress that this approximation should be considered as valid only in the weak anomaly regime and when the fields are not strongly helical. In the strong anomaly regime,
where the chiral anomaly effect strongly influences the evolution of magnetic fields, the characteristic time scale should not be determined by  the fluid-response time -- based on a purely MHD reasoning --
but on the anomaly growth time scale. Moreover, maximally helical modes make no contribution to the Lorentz force and therefore the second order and viscous term in the Navier-Stokes equation
will then become dominant, making this approximation invalid. 
Using the fluid-response time per unit density $\zeta_L$ \cite{Sigltime}, we simplify the velocity field to $\mathbf{v}\approx\zeta_L \mathbf{F}_L$.  
Based on the scaling properties of the induction equation \cite{camp}, the drag time can be taken as $\zeta_L \approx \Gamma(0)[\tau_{\rm eddy}(0) + \gamma t]$, where $\tau_{\rm eddy}(0)$
is the initial eddy turnover time associated with a specific scale $l$, $\tau_{\rm eddy}=l/v$ and $\gamma$ is a constant. 
We concentrate on the case of magnetic fields with fractional helicity, i.e., where the initial helicity is a fraction of maximal helicity, and therefore we have
$h^{\rm bg}_k(0)= \epsilon \rho^{\rm bg}_k(0)/(2k)$, with $0 \leq \epsilon \leq 1$. We model the background magnetic field using an appropriate ansatz that describes the inverse cascade evolution of magnetic energy and helicity \cite{camp}
\begin{eqnarray}
\rho^{\rm bg}_k(t)=\rho^{\rm bg}_k(0) e^{-2k^{2} l_{\rm diss}^{2}} [\cosh(2k l_{\alpha})+\epsilon \sinh(2kl_{\alpha})],\,
\label{background1}
\end{eqnarray}
\begin{eqnarray}
h^{\rm bg}_k(t)=&&h^{\rm bg}_k(0) e^{-2k^{2} l_{\rm diss}^{2}}[\sinh(2k l_{\alpha})+\epsilon \cosh(2kl_{\alpha})],\,
\label{background2}
\end{eqnarray}
where we introduced $\eta_{\rm eff}= (\sigma)^{-1} + 4 \rho^{\rm bg} \zeta_{L}/3$, $\alpha_{B}=- \dot{h}^{\rm bg} \sigma \zeta_{L}/ 3$, $ l_{\rm diss}^{2}= \int_0^\tau d\tau \eta_{\rm eff}$ and $l_{\alpha}= \int_0^\tau d\tau \alpha_{B}$. Note that our background solution for the weak anomaly regime exhibits an inverse cascade for non-vanishing helicity, as expected. Namely, as confirmed by numerous MHD simulations \cite{shiva, brandenburgnew}, helical magnetic fields show the evolution of magnetic spectral modes towards larger scales with time (visible in Fig.~\ref{pic:3hels}). In contrast, the solutions considered in Ref.~\cite{Dvornikov-Semikoz} lead only to resistive decay, with no energy transfer from small to larger scales, and therefore do not describe the effect of inverse cascade.

As pointed out in Ref.~\cite{camp}, the evolution of this background helical field will undergo two different regimes: first, a resistive damping, in which modes of larger wavenumbers decay faster, and thereupon an inverse cascade. In the following we focus on the influence of chiral effects in the latter, due to it being more physically significant. 

The chiral anomaly can have different impacts on the field evolution depending on the evolution of the chiral chemical potential. We consider three limiting cases that illustrate it, namely: when the chirality flipping rates reach an equilibrium with the rate of the source of asymmetry, $\Gamma_f\mu_{R,5} \approx \Pi_{sr}$; when chirality flips dominate, $\Gamma_f\mu_{R,5} \gg \Pi_{sr}$; and when the source rate dominates over chirality flips $\Gamma_f\mu_{R,5} \ll \Pi_{sr}$.      

Starting from \eqref{background1} and \eqref{background2} in these different regimes, shown in Fig.~\ref{pic:3hels} as dashed curves, we present the solutions obtained when adding to it the 
anomaly induced magnetic energy and helicity, \eqref{pert1} and \eqref{pert2}, shown in Fig.~\ref{pic:3hels} as solid curves. All quantities are given in dimensionless units by being scaled to the initial eddy turnover time $\tau_{eddy}(0)\equiv \tau_0$, making our analysis suitable to any particular system.
We have chosen an inverse cascade magnetic energy background spectrum, as described in \S\ref{sec:invcasc},  an initially vanishing anomalous helicity $h^\mu_k(0)=0$.
We take $\Gamma(0) = 0.1$, $\gamma=0.1$ and $\epsilon=7\times10^{-3}$, computed through \eqref{eq:hinduced}.
In order for $\rho_k^{\rm bg}$ and $\rho_k^{\mu}$ to be more easily compared we took different initial asymmetry values $\mu_5(0)$. 

We observe in Fig.~\ref{pic:3hels} that, since we treat the chiral asymmetry as a perturbation,
 the fiducial background magnetic energy remains relatively dominant in the regimes where chirality flips are comparable or dominant with respect to an anomaly source. The initial $\mu_5^0=-10^3\tau_0^{-1}$ was chosen consistent with this assumption and in order to display  its influence on the total magnetic field. The contribution from the anomaly develops into a significant magnetic energy $\rho_k^\mu$ only for timescales $t> 10 \tau_0$ for all studied regimes.  After $\rho_k^\mu$ gets established, $\mu_5$ causes the system to convert the energy stored in the chiral asymmetry to the magnetic field.
In the first case, the change of $\mu_5$ in time is negligible in the inverse
cascade regime. This is characteristic of different stages that occur in various systems that evolve according to chiral MHD, for example around the electroweak phase transition and in the initial stage of evolution of $\mu_5$ in the core of a neutron star (see, e.g. Refs.\cite{si,pa}). In the second case, we take the $\Gamma_{f}(t)=(G_F^2m_e/3t)^2\tau_0^{-1}$, where $G_F$ is the Fermi coupling and $m_e$ is the electron mass, which represents the reaction rate given by the weak interaction, dominant, for instance, shortly after the electroweak crossover \cite{pa}.
 When there is no active source term, this energy is solely drawn from the initial $\mu_5$. 
 We expect that in systems where chirality flips  have a stronger time dependence than in the example taken, $\rho_k^\mu$ would decay faster and no significant traces of the impact of the anomaly would be left. 
In both cases, the total magnetic energy is dominated by the background contribution. On the other hand, when a constant source term, which in Fig.~\ref{pic:3hels} was taken as $\Pi_{sr}=10^{-2}\tau_0^{-1}$, is present and chirality flips absent, $\mu_5$ will tend 
to grow in time in the inverse cascade regime. This can be the case when chirality flips are negligible compared to the source of the anomaly, which can for instance be caused by reactions such as electron capture. After enough time has passed for $\rho_k^\mu$ to become comparable with $\rho_k^{\rm bg}$, the anomalous magnetic energy is going to be dominant, as the lower panel shows, and, therefore, the system exits the weak anomaly regime 
where the presented treatment is valid.

\begin{figure}
\includegraphics[width=0.42\textwidth]{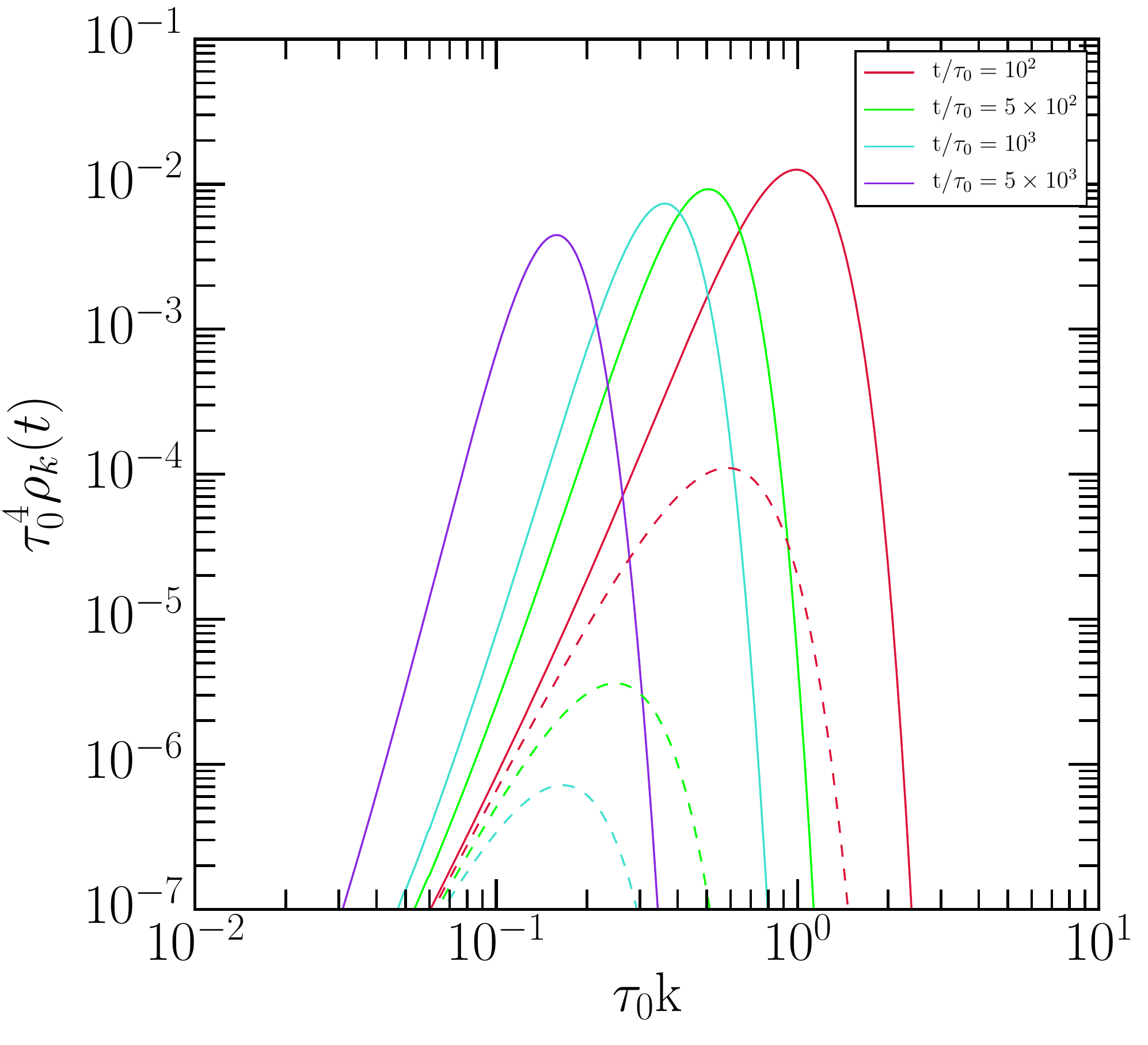}

\caption{\label{pic:seldec} Total magnetic energy density spectrum for an initially vanishing helicity. Dashed lines, computed from \eqref{background1} in the absence of $\mu_5$, showing resistive damping of modes in time. Solid lines in the presence of $\mu_5^0=-10^3\tau_0^{-1}$ for $\Gamma_f\mu_5 = \Pi_{sr}$, which induces a finite helicity, showing an inverse cascade in time.}
\end{figure}

The most dramatical example of the influence of the chiral anomaly in MHD in the weak anomaly regime is the case of no initial helicity in the background magnetic field. Then, equation \eqref{pert2} will give the total helicity at early times, which will enter into \eqref{background1} and \eqref{background2}, and modify the global evolution of magnetic fields. On the other hand, if the anomaly effects on turbulence are not taken into account, then the initially non-helical field will remain non-helical and its evolution described by  \eqref{background1} and \eqref{background2} will lead just to a resistive damping and not to an inverse cascade.  This difference in evolution
related to the anomaly effect is shown in Fig.~\ref{pic:seldec}.

We have thus shown that even in the weak anomaly regime the chiral anomaly effect can have a very important influence on the development of MHD turbulence in two special cases. If there is a 
source term in the system, such that $\Pi_{sr}>\Gamma_{f}\mu_5$, the evolution of the chiral chemical potential will start to dominate the evolution of turbulence, leading to a fast growth
of magnetic energy and then exiting off the weak anomaly regime, as depicted in Fig.~\ref{pic:3hels}. Further analysis of the strong anomaly regime would require a completely new analytical framework for
its description, preferably combined with advanced MHD numerical simulations, and we leave this analysis for the further work.
On the other hand, even in the weak anomaly regime the chiral anomaly effect can lead to the establishment of an inverse cascade in the
case that it did not exist initially. If the fields
are initially non-helical and $\mu_{5}=0$, then the solutions of \eqref{background1} and \eqref{background2} will lead only to resistive damping, with no energy transport from smaller to larger scales, as depicted in
Fig~\ref{pic:seldec}. In the chiral anomaly MHD case, the induced helicity will play the role of an approximately conserved quantity and support the development of the inverse cascade. 

In further studies of MHD turbulence at high temperatures, characteristic for the electroweak scale, it would also be interesting to consider
the anomalous influence on the kinetic helicity and the related chiral vortical effect \cite{Kharzeev:2007tn,Son:2009tf,Jiang:2015cva,sunko}. Here we focused our discussion
on the issue of magnetic helicity in the chiral MHD regime, since it influences the existence of inverse cascades -- which was of central interest
in our work. Moreover, while the chiral anomaly effect is proportional to $\mu_{5}$, the vortical effect is proportional to the square
of the anomalous potential \cite{Kharzeev:2015znc}. Since in realistic cosmological scenarios $\mu_{5}/T \ll 1$ the anomaly effect will typically be dominant compared to the vortical effect.

\section{Conclusion}

Previous studies of the chiral anomaly effect, as well as hypermagnetic fields characterized by the anomalous coupling, have mostly ignored the role of turbulence. Apart from the interesting interplay between velocity, (hyper)magnetic fields and the particle content of the theory, we have discussed how the anomalous modified MHD equations in the turbulent regime can lead to a significantly different time evolution of magnetic fields. We have thus showed in this work that for high enough temperatures -- characteristic, for example, in the early Universe and in proto-neutron stars -- a full description of the considered systems should be given by the chiral MHD turbulence. \\
Focusing on the case of an incompressible fluid in the resistive approximation, we analysed the equations for magnetic and velocity 
field evolution, taking into account the chiral current contributions. With special interest, we considered how chiral modifications 
influence the establishment of an inverse cascade. Creating maximally-helical magnetic fields from initially non-helical configurations, chiral effects can strongly support an
inverse cascade. When anomaly effects are small compared to the standard MHD terms, this manifests as a slower decrease of the magnetic field with time, $\rho_{m} \sim t^{-2/3}$, and as a faster growth of the correlation length, $ \xi_{m} \sim t^{2/3}$, when compared to the evolution of initially non-helical fields predicted by the standard MHD description
(i.e. $\mu_{5}=0$).\\
We then focused on the regime where anomaly effects can not be neglected after helicity was induced. Analysing the evolution of magnetic energy and correlation length in the inertial interval, using a Kolmogorov-like reasoning,
we obtained their scaling with time. If $\mu_{5} \sim t^{n}$ we have $\rho_{m} \sim t^{2(n-1)/3}$ and $\xi_{m} \sim t^{(n+2)/3} $. Taking the special case of an attractor solution, $n=-1/2$, we independently confirm the scalings recently proposed in \cite{yam}. \\
We then considered the weak anomaly regime in more detail. Assuming
that the overall dynamics is determined basically by the standard MHD, we treat the anomaly contribution to magnetic energy and helicity as a small
perturbation to the standard MHD background. Ignoring all the terms higher than the first order in perturbation, we obtained general 
analytical solutions for the anomaly induced helicity and magnetic energy. Using the analytical approximation for the background fields as previously proposed 
in the literature \cite{camp}, we obtained specific solutions for the weak anomaly chiral MHD turbulence in the inverse cascade regime. 
The obtained solutions demonstrate how chiral effects support the inverse cascade and the growth of the correlation length in this regime. The details 
of such an evolution significantly depend on the scaling of the chiral asymmetry potential, $\mu_{5}$, with time -- which is determined by the relationship
between source and chirality-flipping terms. In the case of $\mu_{5}$ growing with time, the induced magnetic energy and helicity
also grow until the assumption of the weak anomaly regime is no longer valid.\\
Thus, the creation of a small amount of magnetic helicity, even if it corresponds to a very
     small change in energy, can lead nevertheless to a considerable change in the evolution of the magnetic field power spectrum due to inverse cascades. In this sense, the final field strength at large scales most interesting
     phenomenologically can be dramatically stronger, although energetically the chiral effect can consistently be treated as a perturbation, as we did in this work.\\
The chiral MHD turbulence description leads to important differences in the evolution of magnetic fields and  chiral asymmetry, with respect to both  
 standard MHD turbulence and the anomaly studies where turbulence effects are ignored. The enhanced growth of the correlation length and the suppressed
decay of magnetic energy that come as a result of the interplay between turbulence and anomaly effects could thus have important consequences 
for different systems. Therefore, the chiral MHD turbulence description could be relevant for
our understanding of different open questions, such as the evolution of cosmic magnetic fields, the baryon asymmetry of the Universe and the creation of magnetic
fields in magnetars.

\begin{acknowledgments}
This work was supported by the ''Helmholtz Alliance for Astroparticle Physics (HAP)'' funded by the Initiative and Networking Fund of the Helmholtz Association and by the Deutsche Forschungsgemeinschaft (DFG) through the Collaborative Research Centre SFB 676 ''Particles, Strings and the Early Universe''.
N. L. thanks Andrej Dundovi\'c for his assistance with numerical computations.

\end{acknowledgments}

\bibliography{apssamp}

\end{document}